\documentclass[10pt]{article}
\usepackage{amssymb}
\usepackage[all, knot]{xy}
\usepackage[all, knot]{}
\usepackage{xcolor}
\usepackage{fourier}
\usepackage{mathtools}
\xyoption{frame}

\begin{document}

\title{Big Bang and Topology}



\author{Torsten Asselmeyer-Maluga\footnote{German Aerospace Center (DLR), Rosa-Luxemburg-Str. 2, 10178, Germany; Email: Torsten.Asselmeyer-Maluga@dlr.de}, 
Jerzy Kr\'ol\footnote{University of Information Technology and Management,  Cognitive Science and Mathematical Modelling Chair, ul. Sucharskiego 2, 35-225 Rzesz\'ow, Poland; Email: jkrol@wsiz.edu.pl} and 
Alissa Wilms\footnote{Freie Universit\"at Berlin, Physics Department,
Arnimallee 14, 14195 Berlin, Germany; Email: alissahelenawilms@gmail.com}}
\maketitle






\begin{abstract}
{In this paper we discuss the initial state of the universe at the
Big Bang. By using ideas of Freedman in the proof of the disk embedding
theorem for 4-manifolds, we describe the corresponding spacetime as
gravitational instanton. The spatial space is a fractal space (wild
embedded 3-sphere). Then we construct the quantum state from this
fractal space. This quantum state is part of the string algebra of
Ocneanu. There is a link to the Jones polynomial and to Witten's topological
field theory. Using this link, we are able to determine the physical
theory (action) to be the Chern-Simons functional. The gauge fixing
of this action determines the foliation of the spacetime and as well
the smoothness properties. Finally, we determine the quantum symmetry
of the quantum state to be the enveloped Lie algebra $U_{q}(sl_{2}(\mathbb{C}))$
where $q$ is the 4th root of unity. 
}
\end{abstract}
\section{Introduction}

What was the initial state of the universe? This question is fundamental
to understand the further development of the universe. However, the
usual extrapolation techniques fail here. Therefore, an answer to
this question seems to be of global nature. As a mathematical method,
therefore, the field of topology would immediately suggest its usefulness here, namely exactly such questions of global nature are answered for spaces within the field.
From this motivation we have therefore awakened the construction of
spacetime by means of topological methods as a starting point. Here,
of course, one can only start with rather general assumptions. All
the more astonishing is the result that only a few quite natural 
assumptions are sufficient to arrive at an unambiguous result.
Exactly the discussion of this approach can be found in the next section
of the paper. Thereby also the part of spacetime can be identified
which belongs to the Big Bang itself. In the search for this part,
topology again plays an important role. Special solutions, gravitational
instantons, represent the tunnel transition to the initial state of
the universe (see \cite{HawHar:83,HawTur:98a}). Finally we get the
rather obvious solution that the 4-disk represents this formation
of the initial state and the 3-sphere is the initial state. However,
one would expect that such state would be a quantum state and not just a \textquotedbl classical\textquotedbl{}
3-manifold. Using our work on quantization by introducing wild embeddings
\cite{AsselmeyerKrol2013}, we simply obtain the quantum state by
the transition to the wildly embedded 3-sphere 
(see Appendix A for a short description of this work). 
{ Here we have to explain the concept of a wild embedding.
In general, an embedding is a map $f:N\to M$ so that $N$ and $f(N)$ are 
topologically equivalent (homeomorphic). The difference between a tame and wild
embedding is given by the description of the image. If the image $f(N)$ can
be described by a finite amount of information (polygons, triangulation etc.)
then the embedding is tame. Examples are the usual knots (as embeddings 
$S^1\to {\mathbb{R}}^3$). In contrast, a wild embedding consist of
an infinite collection of substructures. Examples are the Fox-Artin wild knot
or Alexanders horned sphere. A wild embedding is also given by iterating
a structure like in case of a fractal. In \cite{AsselmeyerKrol2013} we showed that
a wild embedding is a geometric/topological expression for a quantum state.
Therefore we will identify quantum states with wild embeddings and call
it a fractal space. Here we will consider the quantum 3-sphere as a wildly embedded
3-sphere (or a fractal 3-sphere).
}
The description of
the wildly embedded 3-sphere is given in section 3 and its formation
in section 4, using Freedman's idea which he used in solving the disk
embedding problem in dimension 4 \cite{Fre:82,Fre:83,FreQui:90}.
Here the 4-disk is covered by special manifolds (Casson handle)
with a tree-like structure. The description of this structure leads
to the string algebra of Ocneanu \cite{Ocn:88} closely related
to the Jones polynomial \cite{Jon:83,Jon:85} of knot theory. It is
of course a stroke of luck that Witten \cite{Wit:89,Wit:89.2} has
developed a topological quantum field theory exactly for this invariant.
Thus we obtain exactly the physical theory which describes the formation
of the quantum state. The underlying action is the Chern-Simons invariant
and the observable is the Wilson line along the knot. In a special
gauge (axial gauge for the case of a light cone directed into the
future) we obtain a relation to the foliation of spacetime and later
to the Seiberg-Witten theory. These many interrelations to known theories
and approaches show the complexity of the approach. In our forthcoming
work we will turn to the exact description of the initial state and
the implications for the initial distribution of matter and dark matter.

\section{A (coherent) model for the spacetime\label{sec:coherent-model-spacetime}}
In this section, we will describe the model of the spacetime seen
as the space of spacetime events $\mathcal{M}$. At first we start
with three (more or less) obvious assumptions to restrict the class
of spaces $\mathcal{M}$: smooth 4-manifold (we can use the concept
of a differential equation for the dynamics), compactness (every sequence
of events is an event) and simply-connectedness (every time-like loop
can be contracted to maintain causality at least in principle). Then
the spacetime is an open submanifold of $\mathcal{M}$ including examples
like $S^{3}\times\mathbb{R}$. To determine $\mathcal{M}$ completely,
we need the realization of Ricci-flatness in $\mathcal{M}$ representing
the vacuum state (no matter) of general relativity. Together with
the other assumptions, $\mathcal{M}$ is the K3 surface, a Calabi-Yau
space of two complex dimensions (i.e. a 4-dimensional real manifold). In
the following we will discuss the consequences of this approach. In
particular, the K3 surface is a gravitational instanton and using
ideas of Hartle and Hawking, the Big Bang can be understood as tunneling
event induced from a gravitational instanton. We will argue below
that the Big Bang is represented by the 4-disk $D^{4}$ with the initial
state $S^{3}$. Then the corresponding quantum state must be a fractal
space with $S^{3}-$topology. In our previous work, we got a relation
between the quantum state and a so-called wildly embedded 3-sphere
as fractal space. It is the main result of our argumentation in this
section: \emph{the initial state of the universe is the fractal 3-sphere}.
The reader who is willing to accept this assumption can switch to
the next section.

There are infinitely many suitable topologies for the spacetime, seen
as a 4-manifold, and for the space, seen as a 3-manifold. Of course, there
are some heuristics but usually not sufficient for determination of the spacetime uniquely.
Here we will take a different way. Why not trying to determine the space
$\mathcal{M}$ of all possible spacetime-events? Therefore we start
with a definition: let $\mathcal{M}$ be the space of all possible
spacetime events, i.e. the set of all spacetime events carrying a
manifold structure. Then a specific physical system or configuration
is an embedding of a 3-manifold into $\mathcal{M}$ and a dynamics
is an embedding of a cobordism between 3-manifolds (representing the
configuration at the initial and end points) into $\mathcal{M}$. Here,
we assume implicitly that everything can be geometrically/topologically
expressed as submanifolds (see \cite{AsselmeyerRose2012,AsselmeyerBrans2015}).
In the following we will try to discuss this approach and how far
one can go. Some heuristic arguments are rather obvious: 
\begin{enumerate}
\item $\mathcal{M}$ is a smooth 4-manifold, 
\item any sequence of spacetime event has to converge to a spacetime event
and 
\item any loop (time-like or not) must be contracted. 
\end{enumerate}
A dynamics is known to be a mapping of a spacetime event to a new spacetime event.
It is usually a smooth map (differential equations) motivating the first
argument. The second argument expresses the fact that any initial
spacetime event must converge to a final spacetime event. Or, the
limit of any sequence of spacetime events must be converge to a spacetime
event. Then, $\mathcal{M}$ is a compact, smooth 4-manifold. 
The usual or actual spacetime is an open subset of $\mathcal{M}$. 
{ The third argument above is motivated to neglect time-like loops in principle. 
If the underlying spacetime is multiple-connected then there are loops in the
spacetime which cannot be contracted to a point leading to potential time-like loops. 
Therefore, a simple-connected spacetime is a necessary condition to avoid 
closed time-like loops. But, compact spacetime
always admit closed time-like loops, see \cite{HawEll:94}. Therefore, this condition
is not sufficient but the usual (or actual) spacetime is an open subset of $\mathcal{M}$ 
or the usual spacetime is embedded in $\mathcal{M}$. Then, if the usual spacetime
is also simply-connected then because of the non-compactness, see 
\cite{HawEll:94} again, there are no time-like loops.
But to understand the property 'simple-connectedness', we consider a loop in the spacetime. 
If this loop cannot contract then there are two ways or two different curves connecting two different events.
By changing the embedding of the curves via a diffeomorphism (this procedure is called isotopy),
we can deform one curve to agree with other curve. Or, every loop formed by the two curves 
can be contracted. Therefore this argument implies that there are no time-like
loops and the non-compactness of the open subset implies causality.} 
Finally, $\mathcal{M}$ is a compact, simply connected, smooth 4-manifold.

{ The following restrictions of $\mathcal{M}$ will determine the spacetime completely. 
For that reason we demand that the equations
of general relativity are valid without any restrictions. Then the
vacuum equations are given by }
\[
R_{\mu\nu}=0
\]
so that we get Ricci-flatness. But as shown in \cite{Bra:94a,Bra:94b}
and in recent years in \cite{AsselmeyerRose2012,AsselmeyerBrans2015,AsselmeyerMaluga2016},
the coupling to matter can be described by a change of the smoothness
structure. Therefore the modification of the smoothness structure
will produce matter (or sources of gravity). But at the same time,
we need a smoothness structure which can be interpreted as vacuum
given by a Ricci-flat metric. Therefore we will demand that

\begin{enumerate} 
\addtocounter{enumi}{3}
\item$\mathcal{M}$ has to admit a smoothness structure with Ricci-flat
metric representing the vacuum. \end{enumerate}

Interestingly, these four demands are restrictive enough to determine
the topology of $\mathcal{M}$ completely. With the help of Yau's seminal work
\cite{Yau:1978}, that the K3 surface is the unique compact, simply
connected Ricci-flat 4-manifold, we will obtain that\\
 \emph{ $\mathcal{M}$ is topologically equivalent (homeomorphic)
to the K3 surface.}\\
But it is known by the work of LeBrun
\cite{Lebrun96} that there are non-Ricci-flat smoothness structures.
Therefore in a next step, we will determine the smoothness structure of $\mathcal{M}$.
For that purpose, we will present some known results in differential
topology of 4-manifolds (see \cite{AssBra:08} for details and the
construction of the $E_{8}-$manifold): 
\begin{itemize}
\item there is a compact, contractible submanifold $A\subset\mathcal{M}$
(called Akbulut cork) so that cutting out $A$ and reglue it (by an
involution) will produce a new smoothness structure, 
\item $\mathcal{M}$ splits topologically into 
\begin{equation}
|E_{8}\oplus E_{8}|\#\underbrace{\left(S^{2}\times S^{2}\right)\#\left(S^{2}\times S^{2}\right)\#\left(S^{2}\times S^{2}\right)}_{3\left(S^{2}\times S^{2}\right)}=2|E_{8}|\#3(S^{2}\times S^{2})\label{eq:splitting-K3}
\end{equation}
two copies of the $E_{8}-$manifold and three copies of $S^{2}\times S^{2}$
and 
\item the 3-sphere $S^{3}$ is a submanifold of $A$. 
\end{itemize}
In \cite{AsselmeyerKrol2012} we already discussed this case. From
the topological point of view, any sum of $E_{8}-$manifolds and $S^{2}\times S^{2}$
is realized by a closed, simply-connected, topological 4-manifold
but not all topological 4-manifolds are smooth manifolds. To clarify
this point, let us consider the 4-manifold which splits topologically
into $p$ copies of the $|E_{8}|$ manifold and $q$ copies of $S^{2}\times S^{2}$
or 
\[
p|E_{8}|\#q\left(S^{2}\times S^{2}\right)\:.
\]
Then, this 4-manifold is smoothable for every $q$ but $p=0$. The
first combination for $p\not=0$ is the pair of numbers $p=2,q=3$
(which is the K3 surface). Any other combination ($p=2,q<3$ or every
$q$ and $p=1$) is forbidden as shown by Donaldson \cite{Don:83}.
Therefore the simplest combination of $|E_{8}|$ and $S^{2}\times S^{2}$
is realized by the K3 surface.

Now we consider the smooth K3 surface which is Ricci-flat, simply
connected and smooth. A main part in the following discussion will
be the usage of the smoothness condition. As discussed above, the
smoothness structure is determined by the Akbulut cork $A$. Furthermore
as argued above, the smoothness structure is strongly related to the
appearance of matter (see \cite{AsselmeyerRose2012,AsselmeyerBrans2015,AsselmeyerMaluga2016})
and this process is strongly connected to the evolution of our cosmos
(see \cite{AsselmeyerKrol2018a,AsselmeyerKrol2018b}). This process
is known as reheating after the inflationary phase. Therefore, the
Akbulut cork (including its embedding) should represent the inflationary
phase with reheating. We have already discussed this partly in our works (see \cite{AsselmeyerKrol2018c,AsselmeyerMaluga2016} for first results in this direction).

The central submanifold determining the smoothness structure is the
Akbulut cork $A$, a contractible submanifold with boundary $\partial A$.
As shown by Freedman \cite{Fre:82}, the Akbulut cork is build from
a homology 3-sphere which will become the boundary $\partial A$.
The difference to a usual 3-sphere $S^{3}$ is given by the so-called
fundamental group, the equivalence class of closed loops up to deformation
(homotopy) with concatenation as group operation. In principal, one
constructs a cobordism between $S^{3}$ and the homology 3-sphere
$\partial A$. All elements of the fundamental group will be killed
by adding appropriate disks. At the end, one can add a 4-disk to get
the full contractible cork $A$. The topology of $\partial A$ depends
strongly on the topology of $\mathcal{M}$. In case of the K3 surface,
$\partial A$ is known to be a Brieskorn spheres, precisely the 3-manifold
\[
\partial A=\Sigma(2,5,7)=\left\{ x,y,z\in\mathbb{C}\,|\,x^{2}+y^{5}+z^{7}=0\,|x|^{2}+|y|^{2}+|z|^{2}=1\right\} \:.
\]
{ The construction of the smoothness structures is based on the work 
\cite{Biz:95,BizGom:96}. The smoothness structure depends on the Casson handle 
(used to construct an exotic ${\mathbb{R}}^4$ in the cited work). 
A Casson handle is uniquely determined by a branched tree. Then the simplest
Casson handle is given by an unbranched tree and we will choose this smoothness
structure in the following. The corresponding K3 surface is constructed in \cite{BizGom:96}.}

The embedding of the Akbulut cork is essential for the following results.
In \cite{AsselmeyerKrol2018a} it was shown that the embedded cork
admits a hyperbolic geometry if the underlying K3 surface has an exotic
smoothness structure. Additionally the open neighborhood $N(A)$ of
the Akbulut cork in the K3 surface is an exotic $\mathbb{R}^{4}$,
i.e. a space homeomorphic to the Euclidean space $\mathbb{R}^{4}$
but not diffeomorphic to it. In the following we will denote this
exotic $\mathbb{R}^{4}$ as $R^{4}$. One of the characterizing properties
of an exotic $\mathbb{R}^{4}$ (all known examples) is the existence
of a compact subset $K\subset R^{4}$ which cannot be surrounded by
any smoothly embedded 3-sphere (and homology 3-sphere bounding a contractible,
smooth 4-manifold). But there is always a topologically embedded 3-sphere
i.e. this 3-sphere is wildly embedded. In \cite{AsselmeyerMaluga2016}
we described this wildly embedded 3-sphere explicitly (denoted as
$Y_{\infty}$) and we showed in \cite{AsselmeyerKrol2013} that this
wildly embedded 3-sphere can be understood as quantum state i.e. it
is the deformation quantization of a tame (or usual) embedding. The
notation \emph{wildly embedded} or \emph{wild} is purely mathematical.
Instead we will denote this wild 3-sphere as \emph{fractal 3-sphere}.
But at first we will look at the Akbulut cork $A$ which can be decomposed
as
\begin{equation}
A=D^{4}\cup_{S^{3}}W(S^{3},\partial A)\label{eq:Akbulut-cork}
\end{equation}
where $W(S^{3},\partial A)$ describes a cobordism between the 3-sphere
and the boundary $\partial A=\Sigma(2,5,7)$. In \cite{AsselmeyerKrol2018a}
we discussed this cobordism $W(S^{3},\partial A)$ as the first (inflationary)
transition $S^{3}\to\partial A$ from the initial state (the 3-sphere)
to a non-trivial space (containing matter). Then by using the embedding
of $A$ into the K3 surface, we identify the 3-sphere (boundary of
$D^{4}$) with the wild 3-sphere $Y_{\infty}$ (from the open neighborhood
$N(A)$), or the initial state of our model of the universe is a fractal
3-sphere (which is a quantum state, see \cite{AsselmeyerKrol2013,AsselmeyerMaluga2016}).
With this identification in mind, we are able to interpret the first
transition $W(S^{3},\partial A)$ (from the wild 3-sphere to the (classical)
non-trivial state $\partial A$) as decoherence process, see \cite{AsselmeyerKrol2013c}.
In \cite{AsselmeyerKrol2018a}, we discussed a second transition leading
to a cosmological constant. Finally we have the two transitions
\begin{equation}
S^{3}\stackrel{cork}{\longrightarrow}\partial A=\Sigma(2,5,7)\stackrel{gluing}{\longrightarrow}P\#P\label{eq:two-transitions}
\end{equation}
where $P$ denotes the Poincare sphere. In this paper we are interested
in the formation of the initial state (the fractal 3-sphere), also
called Big Bang. Using the decomposition (\ref{eq:Akbulut-cork}),
this formation is expressed in spacetime via the 4-manifold $D^{4}$
having the boundary $\partial D^{4}=S^{3}$, the (fractal) 3-sphere.
Again, the embedding of $D^{4}$ into the K3 surface is important,
otherwise one will never obtain the fractal 3-sphere as boundary.
Therefore, many properties of the K3 surface go over to $D^{4}$ by
using the embedding. 

To describe this embedding, we need the following fact: the K3 surface
is a gravitational instanton. We implicitly used this fact above when
we constructed a simply-connected, Ricci-flat spacetime (uniquely
given by the K3 surface). In general, an instanton is a field configuration,
which is interpreted as a tunneling effect between topologically in-equivalent
sectors of the vacuum. The term \textquotedbl gravitational instanton\textquotedbl{}
is usually used for 4-manifolds whose Weyl tensor is self-dual and
fulfills the Einstein condition $Ric=\Lambda g$. Usually it is assumed
that the metric is asymptotic to the standard metric of Euclidean
4-space. In case of the K3 surface, there is the phenomenon that gravitational
instantons are created by bubbling off a subspace. 
{ Here we recommend the recent publication \cite{Hein2018} for the
description of this process.} To state it more
precisely, there is a family of hyperk\"ahler metrics $g_{\beta}$ on
a K3 surface which collapse to an interval $[0,1]$ in the Gromov-Hausdorff
limit ($\beta\to\infty$ with metrics $dt^{2}$) with Taub-NUT bubbles
in the interior and Tian-Yau metrics at the endpoints. For the embedding
of $D^{4}$ we choose the Taub-NUT metric in the (open) neighborhood
of the boundary. But what about the interior of $D^{4}$? Here, we
have to use the elliptic fibration of the K3 surface (as torus bundle
over the $S^{2}$ with singular fibers, see \cite{GomSti:1999}).
Then we can describe the embedded $D^{4}$ by the Eguchi-Hanson metrics
(a gravitational instanton). This metric is a Riemannian metrics.
Here, the signature of the metrics changes from the Riemannian signature
(for $D^{4}$) to the Lorentzian signature (for $\partial D^{4}\times(0,1)$).
{ In a recent publication \cite{Hein2021}, a gravitational
instanton with these properties is constructed. The construction
used explicitly the hyperk{\"a}hler structure ($SU(2)$ holonomy group).
The gluing of the instanton solutions can be done by using the work \cite{Hein2018}.
}
\par
As explained above, the boundary $\partial D^{4}$ is identified with
the wild (or fractal) 3-sphere. Then the signature change of the metric
can be identified with the formation of this fractal 3-sphere. Here
we follow the usual interpretation (Hartle-Hawking and Hawking-Turok
see \cite{HawHar:83,HawTur:98a}) that the gravitational instanton
$D^{4}$ represents the Big Bang (via a tunneling event) leading to
the quantum state of the universe. In \cite{AsselmeyerKrol2013} we
showed that a quantum state can be topologically understood as a wildly
embedded 3-sphere or a fractal 3-sphere for short. Therefore we will
argue accordingly that the quantum state of the universe (as initial
state) is represented by the fractal 3-sphere. In the next section
we will describe this fractal 3-sphere explicitly.

\section{The construction of the fractal 3-sphere as quantum state}\label{fractal-3-sphere}

In \cite{AsselmeyerKrol2018a,AsselmeyerKrol2018b,AsselmeyerKrol2018c,AsselmeyerMaluga2018d},
we described a model for the cosmic evolution which is in good agreement
to current measurements \cite{PlanckCosmoParameters2013,PlanckInflation2013}.
Amazingly as discussed above, we are able to extrapolate the state
at the Big Bang \cite{AsselmeyerMaluga2016,AsselmeyerMaluga2018d}:
a fractal 3-sphere as boundary of a 4-disk $D^{4}$, i.e. a gravitational
instanton as transition (tunneling) to a fractal 3-sphere representing
the quantum state \cite{AsselmeyerKrol2013}. Furthermore as explained
above, this fractal 3-sphere is part of $R^{4}$, an exotic $\mathbb{R}^{4}$.
{ 
Before we start with the construction of the fractal 3-sphere, 
we will describe the physical ideas behind the construction. 
In the introduction, we explained the concept of a wild embedding (or fractal space). 
In short, a wild embedding is a submanifold (image of an embedding map) which must be 
decomposed into infinitely many substructures (polygons etc.). 
Therefore it contains an infinite amount of information. In our previous work 
we showed that the wild embedding is an expression for a quantized geometry. 
In case of a fractal 3-sphere (as wildly embedded 3-sphere), one decomposes the 3-sphere 
into similar looking pieces with constant curvature. Every piece has a different curvature 
so that the whole fractal 3-sphere represents the set of possible curvatures. 
These structures appear at all scales. Because of this property, we have to 
use the methods of noncommutative geometry to get a rigorous definition of this procedure. 
The following construction of the fractal 3-sphere is directly motivated by the exotic smoothness 
structure. The basic structure is a tree (used to define the Casson handle). Every part of the tree like 
edge or vertex is associated to a 3-manifold. For the whole tree, one gets an infinitely complicated 3-
manifold which is topologically equivalent to a 3-sphere. This fractal 3-sphere is the boundary of a 4-
disk or 4-ball described in the next section and representing the Big Bang as gravitational instanton (via 
a tunneling event).
}\par
In \cite{AsselmeyerMaluga2016} we described this fractal 3-sphere
as a sequence of 3-manifolds 
\[
Y_{0}\to Y_{1}\to\cdots\to Y_{\infty}
\]
with increasing complexity. { At first we want to comment on the uniqueness 
of the construction. The sequence of 3-manifolds is determined by the smoothness
structure or better by the Casson handle which is used to construct this
structure. Every Casson handle is represented by a tree. This tree is translated
into a link: every $n$-branching point (vertex of the tree) is given by a Whitehead 
link with $n$ circles and every line (edge of the tree) is given by the circle 
of the Whitehead link. In the previous section, we introduced the smoothness structure
as given by the unbranched tree. Obviously, the unbranched tree is a subtree for any
other more complex tree. It is a fundamental property of Casson handles (see \cite{Fre:82})
that a Casson handle $CH_1$ embeds into another Casson handle $CH_2$, say $CH_1\subset CH_2$, iff the tree of $CH_2$ embeds into the tree of $CH_1$. Therefore any other Casson handle
embeds into the Casson handle represented by the unbranched tree. This property is unique
for the smoothness structure and the construction of the fractal 3-sphere. }

For completeness we will shortly explain
the construction. The 3-manifold $Y_{0}$ is given by surgery ($0-$framed)
along the pretzel knot $(-3,3,-3)$ (or the knot $9_{46}$ in Rolfson
notation), $Y_{1}$ is constructed by $0-$framed surgery along the
Whitehead double of the pretzel knot $(-3,3,-3)$ and finally $Y_{n}$
is constructed by $0-$framed surgery along the $n$th Whitehead double
of the pretzel knot $(-3,3,-3)$. In the limit $n\to\infty$, we obtained
$Y_{\infty}$ as $0-$framed surgery along the $\infty$th Whitehead
double of the pretzel knot $(-3,3,-3)$ (a so-called wild knot). This
3-manifold $Y_{\infty}$ is the fractal 3-sphere (it has the topology
of a 3-sphere by a theorem of Freedman \cite{Fre:82}). The whole
process can be seen as an iteration process at the level of 3-manifolds:
we start with $Y_{0}$ and end with $Y_{\infty}$, the fractal 3-sphere.

To understand this abstract construction (via Dehn surgery or Kirby
calculus \cite{GomSti:1999}) we have to describe the construction
of the first 3-manifold $Y_{0}$ more carefully. For that purpose,
we have to describe Dehn surgery or surgery along a knot. If we remove
a thicken knot $N(K)=K\times D^{2}$ (so-called tubular neighborhood)
from the 3-sphere $S^{3}$ then one obtains the knot complement $C(K)=S^{3}\setminus N(K)$.
Now we glue in one solid torus $D^{2}\times S^{1}$ to $C(K)$ by
a mapping of the boundary $\phi:\partial C(K)=T^{2}\to\partial(D^{2}\times S^{1})=T^{2}$
so that we get 
\[
M_{K,\phi}=C(K)\cup_{\phi}\left(D^{2}\times S^{1}\right)\:.
\]
All closed curves on a torus can be generated by the two possible
non-contracting curves $m,\ell$ the meridian and longitude, respectively.
In principle, any closed curve $\gamma$ on a torus $T^{2}$ is given
by two numbers with ${\displaystyle [\gamma]=[a\ell+bm]}$ (for the
homotopy classes). Then the map $\phi$ is characterized by a mapping
of the meridian $m$ of one torus to the curve $\gamma$ determined
by the ratio $r=b/a$ (including $\infty$ for $a=0$) called the
frame number. As a warm-up example, we consider the 0-framed surgery
along the unknot $S^{1}$ in $S^{3}$. The knot complement of the
unknot $C(S^{1})=D^{2}\times S^{1}$ is glued to another solid torus
$D^{2}\times S^{1}$ (along its boundary $\partial(D^{2}\times S^{1})=S^{1}\times S^{1}$)
with framing $0$ which means that the meridian of $\partial C(S^{1})$
is mapped to the meridian of $\partial(D^{2}\times S^{1})$. But that
means that $D^{2}\times S^{1}$ is glued to $D^{2}\times S^{1}$ along
the boundary, i.e. $(D^{2}\cup_{\partial D^{2}}D^{2})\times S^{1}=S^{2}\times S^{1}$.
Therefore $0-$framed surgery along the unknot gives $S^{2}\times S^{1}$.
Interestingly, $0-$framed surgery along any knot produces a 3-manifold
which is very similar to $S^{2}\times S^{1}$ (having the same homology).
Every $Y_{n}$ in the sequence above is produced by $0-$framed surgery
along an knot of increasing complexity. One starts for $n=0$ with
the knot $9_{46}$ (in Rolfson notation) producing $Y_{0}$ then $n=1$
with $Y_{1}$ is produced by the Whitehead double $Wh_{1}(9_{46})$
of this knot, $Y_{2}$ is given by the second iterated Whitehead double
$Wh_{2}(9_{46})$ and so on. In the limit $n\to\infty$ one gets $Y_{\infty}$
as $0-$framed surgery along the $\infty-$iterated Whitehead double
$Wh_{\infty}(9_{46})$ of $9_{46}$ (a so-called wild knot). But this
limit changes the topology of $Y_{\infty}$. For every finite $n\geq0$,
$Y_{n}$ has the same homology as $S^{2}\times S^{1}$ but $Y_{\infty}$
is topologically equivalent to $S^{3}$ (by a theorem of Freedman
\cite{Fre:82}). 

In \cite{AsselmeyerKrol2013,AsselmeyerMaluga2016} we constructed
a quantum state from a wild embedding. Main idea is the description
of the wild embedding by using operator algebras in the spirit of
noncommutative geometry. This relation is strict: the wild embedding
is in one-to-one relation to a foliation with leaf space a factor
$I\!I\!I$ von Neumann algebra known as the observable algebra of
a quantum field theory. To understand this relation from a geometrical
point of view, we will use the decomposition of the factor $I\!I\!I$
into a factor $I\!I$ and a one-parameter group of automorphisms.
We remark that this decomposition was used by Rovelli and Connes \cite{ConRov:94}
to introduce a time variable in quantum gravity. This decomposition
means that in some sense the intractable factor $I\!I\!I$ can be
reduced to the easier accessible factor $I\!I$ (operators of finite
trace). 

For completeness we will also present the construction (see \cite{AsselmeyerKrol2013})
of the $C^{*}-$algebra from the wild embedded 3-sphere. Let $I:S^{3}\to\mathbb{R}^{4}$
be a wild embedding of codimension-one so that $I(S^{3})=S_{\infty}^{3}=Y_{\mathcal{T}}$.
Now we consider the complement $\mathbb{R}^{4}\setminus I(S^{3})$
which is non-trivial, i.e. $\pi_{1}(\mathbb{R}^{4}\setminus I(S^{3}))=\pi\not=1$.
Now we define the $C^{*}-$algebra $C^{*}(\mathcal{G},\pi$) associated
to the complement $\mathcal{G}=\mathbb{R}^{4}\setminus I(S^{3})$
with group $\pi=\pi_{1}(\mathcal{G})$. If $\pi$ is non-trivial then
this group is not finitely generated. From an abstract point of view,
we have a decomposition of $\mathcal{G}$ by an infinite union
\[
\mathcal{G}=\bigcup_{i=0}^{\infty}C_{i}
\]
of 'level sets' $C_{i}$. Then every element $\gamma\in\pi$ lies
(up to homotopy) in a finite union of levels. 

The basic elements of the $C^{*}-$algebra $C^{*}(\mathcal{G},\pi$)
are smooth half-densities with compact supports on $\mathcal{G}$,
$f\in C_{c}^{\infty}(\mathcal{G},\Omega^{1/2})$, where $\Omega_{\gamma}^{1/2}$
for $\gamma\in\pi$ is the one-dimensional complex vector space of
maps from the exterior power $\Lambda^{k}L$ ($\dim L=k$), of the
union of levels $L$ representing $\gamma$, to $\mathbb{C}$ such
that 
\[
\rho(\lambda\nu)=|\lambda|^{1/2}\rho(\nu)\qquad\forall\nu\in\Lambda^{2}L,\lambda\in\mathbb{R}\:.
\]
For $f,g\in C_{c}^{\infty}(\mathcal{G},\Omega^{1/2})$, the convolution
product $f*g$ is given by the equality
\[
(f*g)(\gamma)=\intop_{\gamma_{1}\circ\gamma_{2}=\gamma}f(\gamma_{1})g(\gamma_{2})
\]
with the group operation $\gamma_{1}\circ\gamma_{2}$ in $\pi$. Then
we define via $f^{*}(\gamma)=\overline{f(\gamma^{-1})}$ a $*$operation
making $C_{c}^{\infty}(\mathcal{G},\Omega^{1/2})$ into a $*$algebra.
Each level set $C_{i}$ consists of simple pieces (in case of Alexanders
horned sphere, we will explain it below) denoted by $T$. For these
pieces, one has a natural representation of $C_{c}^{\infty}(\mathcal{G},\Omega^{1/2})$
on the $L^{2}$ space over $T$. Then one defines the representation
\[
(\pi_{x}(f)\xi)(\gamma)=\intop_{\gamma_{1}\circ\gamma_{2}=\gamma}f(\gamma_{1})\xi(\gamma_{2})\qquad\forall\xi\in L^{2}(T),\forall x\in\gamma.
\]
The completion of $C_{c}^{\infty}(\mathcal{G},\Omega^{1/2})$ with
respect to the norm 
\[
||f||=\sup_{x\in\mathcal{G}}||\pi_{x}(f)||
\]
makes it into a $C^{*}$algebra $C_{c}^{\infty}(\mathcal{G},\pi$).
Finally we are able to define the $C^{*}-$algebra associated to the
wild embedding. Using a result in \cite{AsselmeyerKrol2013}, one
can show that the corresponding von Neumann algebra is the factor
$I\!I\!I_{1}$. This algebra is the observable algebra of a free (algebraic)
quantum field theory with one vacuum vector \cite{Borchers2000}.
Here we will discuss an alternative way to construct the factor $I\!I\!I_{1}$.
For that purpose, we look again at the construction of the wild 3-sphere
$Y_{\infty}$. The $\infty-$iterated Whitehead double $Wh_{\infty}(9_{46})$
of the knot $9_{46}$ gives a wild knot $\mathcal{K}$ and $Y_{\infty}$
can be constructed by
\[
Y_{\infty}=C(\mathcal{K})\cup\left(D^{2}\times S^{1}\right)
\]
the $0-$framed surgery. In \cite{AsselmeyerKrol2013}, we discussed
the known result that the (deformation) quantization of the geometric
structures (space of constant curvature) is given by the Kauffman
bracket skein module. For $Y_{\infty}$ it means that we have to consider
the Kauffman bracket skein module $K_{h}(C(\mathcal{K}))$ of $C(\mathcal{K})$.
Here, it is known that $K_{h}(C(\mathcal{K}))$ is a module over the
noncommutative torus which is related (for $h=0)$ to the boundary
$\partial C(\mathcal{K})=T^{2}$. The noncommutative torus defines
a factor $I\!I_{\infty}$ algebra and we will show in our forthcoming
work that the whole $K_{h}(C(\mathcal{K}))$ gives the factor $I\!I\!I_{1}$. 

\section{The quantum spacetime at the Big Bang}

In section \ref{sec:coherent-model-spacetime} we described the Big
Bang as gravitational instanton $D^{4}$ (induced from spacetime,
the K3 gravitational instanton). The initial state of the universe
is given as the boundary $\partial D^{4}=S^{3}$, a wild 3-sphere,
via a tunneling process (Hartle-Hawking). Usually, nothing is known
about the formation of the initial state via the tunneling process.
In contrast, we have here the comfortable situation that there is
a relation between the boundary - the wild 3-sphere - and the interior
of the 4-disk. There is a process for the formation of the wild 3-sphere,
which is divided into an infinite number of subprocesses, called Casson
handles. This structure is called the design and was developed for the
classification of 4-manifolds \cite{Fre:82,FreQui:90}. All subprocesses
can be parameterized by all paths in a binary tree. The detailed construction
of these Casson handles is unimportant for the following (but see
\cite{Fre:82}). 
{ 
Again before we start with the construction, we will discuss the physics behind it. Like in case of the 
fractal 3-sphere, the design is a geometric/topological expression for the quantum state of the spacetime. 
Here, it is the formation of the fractal 3-sphere seen as the boundary of the 4-disk. The design is a 
summation over all possible formation processes. It is an expression for the functional integral. Like for 
the construction of the fractal 3-sphere, we also get complicated substructures at all scales so that we 
need the methods of the noncommutative geometry again. Here, the formation process is parametrized by a 
binary tree where every path is a particular process. But we need all processes or pathes of the binary. 
Therefore we associate to every path an operator which consists of a sum of elementary operators 
(projection operators). Then one gets directly an operator algebra (Temperley-Lieb algebra) which can be 
interpreted as an algebra of field operators. Here, we use the fact that we consider paths of a binary 
tree: the operator algebra is the algebra of fermion field operators. Interestingly, the expectation value 
in this algebra is related to a structure (Jones polynomial) which is well-known for 3-dimensional 
manifolds and knots. Now we argue backwards: the expectation value is defined by a functional integral 
with Chern-Simons action in agreement with our previous work. The Chern-Simons action in the light cone 
gauge is interpreted as invariant of the underlying foliation of the spacetime. Again with the help of 
noncommutative geometry, we are able to get a kind of quantum action (the so-called flow of weigths). We 
remark that at the topological level we have a kind of duality between the design (4D) and links (3D) 
which will be further investigated in our forthcoming work.
}\par
The design $S(Q)$ is a structure to label all Casson handle which
embed in a given Casson handle $Q$. 
{ In our case, this Casson handle $Q$ is represented by an
unbranched tree.} Then this Casson handle $Q$
represents (in some sense) all Casson handles. We will define this
design $S(Q)$ to be the quantum state of $Q$. Below we will determine
the operator algebra associated to $Q$ and we will show that this
algebra is a von Neumann algebra of finite trace as well with one vacuum
vector (factor $I\!I_{1}$). But at first we will describe the construction
of the design $S(Q)$. { In \cite{AsselmeyerKrol2010} we also described this 
construction but in a different context. For completeness we will present this
construction again.}

According to Freedman (\cite{Fre:82} p.393), a Casson handle is represented
by a labeled finitely-branching tree $Q$ with basepoint $\star$,
having all edge paths infinitely extendable away from $\star$. Each
edge should be given a label $+$ or $-$ . The tree $Q$ is fixed
generating the wild 3-sphere (as the boundary of $D^{4}$). Then Freedman
(\cite{Fre:82} p.398) constructs another labeled tree $S(Q)$ from
the tree $Q$. There is a base point from which a single edge (called
``decimal point'') emerges. The tree is binary: one edge enters
and two edges leaving a vertex. The edges are named by initial segments
of infinite base 3-decimals representing numbers in the standard ``middle
third'' Cantor set $C.s.\subset[0,1]$. This kind of Cantor set is
given by the following construction: Start with the unit Interval
$S_{0}=[0,1]$ and remove from that set the middle third and set $S_{1}=S_{0}\setminus(1/3,2/3)$
Continue in this fashion, where $S_{n+1}=S_{n}\setminus\{\mbox{middle thirds of subintervals of \ensuremath{S_{n}}}\}$.
Then the Cantor set $C.s.$ is defined as $C.s.=\cap_{n}S_{n}$. With
other words, if we using a ternary system (a number system with base
3), then we can write the Cantor set as $C.s.=\{x:x=(0.a_{1}a_{2}a_{3}\ldots)\mbox{ where each \ensuremath{a_{i}=0} or \ensuremath{2}}\}$.
Each edge $e$ of $S(Q)$ carries a label $\tau_{e}$ where $\tau_{e}$
is an ordered finite disjoint union of 6-level-subtrees. There is
three constraints on the labels which leads to the correspondence
between the $\pm$ labeled tree $Q$ and the (associated) $\tau$-labeled
tree $S(Q)$.

Every path in $S(Q)$ represents one tree leading to a Casson handle.
Any subtree represents a Casson-handle which embeds in $Q$, see above. Now we
will introduce an (operator) algebra structure on $S(Q)$. For that
purpose, we have to consider pairs of paths in the (dual) tree of
$S(Q)$. Thus we have to concentrate on the so-called string algebra
according to Ocneanu \cite{Ocn:88}. For that purpose we define a
non-negative function $\mu:Edges\to{\mathbb{C}}$ together with the
adjacency matrix $\triangle$ acting on $\mu$ by 
\[
\triangle\mu(x)=\sum\limits _{{v\in Edges\atop {s(v)=x\atop r(v)=y}}}\mu(y)
\]
where $s(v)$ and $r(v)$ denote the source and the range of an edge
$v$. A path in the tree is a succession of edges $\xi=(v_{1},v_{2},\ldots,v_{n})$
where $r(v_{i})=s(v_{i+1})$ and we write $\tilde{v}$ for the edge
$v$ with the reversed orientation. Then, a string on the tree is
a pair of paths $\rho=(\rho_{+},\rho_{-})$, with $s(\rho_{+})=s(\rho_{-})$,
$r(\rho_{+})\sim r(\rho_{-})$ which means that $r(\rho_{+})$ and
$r(\rho_{-})$ ending on the same level in the tree and $\rho_{+},\rho_{-}$
have equal lengths i.e. $|\rho_{+}|=|\rho_{-}|$ expressing the previous
described property $r(\rho_{+})\sim r(\rho_{-})$ too. Now we define
an algebra $String^{(n)}$ with the linear basis of the $n$-strings,
i.e. strings with length $n$ and the additional operations: 
\begin{eqnarray*}
(\rho_{+},\rho_{-})\cdot(\eta_{+},\eta_{-}) & = & \delta_{\rho_{-},\eta_{+}}(\rho_{+},\eta_{-})\\
(\rho_{+},\rho_{-})^{*} & = & (\rho_{-},\rho_{+})
\end{eqnarray*}
where $\cdot$ can be seen as the concatenation of paths. We normalize
the function $\mu$ by $\mu(root)=1$. Now we choose a function $\mu$
in such a manner that 
\begin{equation}
\triangle\mu=\beta\mu\label{graph-laplace}
\end{equation}
for a complex number $\beta$. Then we can construct elements $e_{n}$
in the algebra $String^{(n+1)}$ by 
\begin{equation}
e_{n}=\sum\limits _{{|\alpha|=n-1\atop |v|=|w|=1}}\frac{\sqrt{\mu(r(v))\mu(r(w))}}{\mu(r(\alpha))}(\alpha\cdot v\cdot\tilde{v},\alpha\cdot w\cdot\tilde{w})\label{Jones-projection}
\end{equation}
which are the generators of the so-called Temperley-Lieb algebra.
A \emph{Temperley-Lieb algebra} is an algebra with unit element $\mathbf{1}$
over a number field $K$ generated by a countable set of generators
$\{e_{1},e_{2},\ldots\}$ with the defining relations
\begin{eqnarray}
e_{i}^{2}=\tau\cdot e_{i}\,, & e_{i}e_{j}=e_{j}e_{i}\,:\,|i-j|>1,\nonumber \\
e_{i}e_{i+1}e_{i}=\tau e_{i}\,, & e_{i+1}e_{i}e_{i+1}=\tau e_{i+1}\,,\,e_{i}^{*}=e_{i}\label{Jones-algebra}
\end{eqnarray}
where $\tau$ is a real number in $(0,1]$. By \cite{Jon:83}, the
Temperley-Lieb algebra has a uniquely defined trace $Tr$ which is
normalized to lie in the interval $[0,1]$. The generators (\ref{Jones-projection})
also fulfill these algebraic relations (\ref{Jones-algebra}) where
$\tau=\beta^{-2}$. The trace of the string algebra given by 
\begin{equation}
tr(\rho)=\delta_{\rho_{+},\rho_{-}}\beta^{-|\rho|}\mu(r(\rho))\label{eq:trace-Jones-poly}
\end{equation}
and defines on $A_{\infty}=(\bigcup\limits _{n}String^{(n)},tr)$
an inner product by $\langle x,y\rangle=tr(xy^{*})$ giving after
completion the Hilbert space $L^{2}(A_{\infty},tr)$. 

Now we will determine the parameter $\tau$. Originally, Ocneanu introduce
its string algebra to classify the splittings of modules over an operator
algebra (see also \cite{GoHaJo:89}). Thus, to determine this parameter
we look for the simplest generating structure in the tree. The simplest
structure in the binary tree $S(Q)$ is one edge which is connected
with two other edges. This graph is represented by the following adjacency
matrix 
\begin{eqnarray*}
\left(\begin{array}{ccc}
0 & 1 & 1\\
1 & 0 & 0\\
1 & 0 & 0
\end{array}\right)
\end{eqnarray*}
having eigenvalues $0,\sqrt{2},-\sqrt{2}$. According to our definition
above, $\beta$ is given by the greatest eigenvalue of this adjacency
matrix, i.e. $\beta=\sqrt{2}$ and thus $\tau=\beta^{-2}=\frac{1}{2}$.
Then, without proof, we state that the algebra $R$ is given by the
Clifford algebra on ${\mathbb{R}}^{\infty}$. The coefficients of
this algebra are given by a map $\mu:Edges\to{\mathbb{C}}$.

The definition of the trace (\ref{eq:trace-Jones-poly}) (or better
the inner product) has a strong link to knot theory. This algebra
(\ref{Jones-algebra}) was used by Jones \cite{Jon:83,Jon:85} to
define a new knot invariant. Therefore, we can interpret every expectation
value as the knot/link invariant of a certain knot/link (represented
by a braid) or a sum of these invariants. But before we have to map
the projectors $e_{i}$ to the generators $g_{i}$ so that $e_{k}=\frac{1}{1+i}+\frac{1}{\sqrt{2}i}g_{k}$
(for the special value $\tau=\frac{1}{2}$), see \cite{Jon:85}. Then
every generator $b_{i}$ of the braid group $B_{n}$ is mapped to
$g_{i}$ (and vice versa). So, the expectation value is associated
to a (formal sum) of braids. The closure of these braids are links
or every string $\rho$ defines a (formal) sum of links $L_{\rho}$.
Then $tr(\rho)$ must be equal (by definition) to the Jones polynomial
$V_{L_{\rho}}(t=i)$ for the link $L_{\rho}$ for the special value
$t=i$ (in general $t=\exp(i\pi\tau)$). The value of the Jones polynomial
for $t=i$ is known to be
\[
V_{L_{\rho}}(i)=-\left(\sqrt{2}\right)^{\ell-1}(-1)^{Arf(L_{\rho})}
\]
where $\ell$ is the number of components for $L_{\rho}$ and $Arf(L)$
is the Arf-invariant of the link (see \cite{Mur:86} for the proof
of the result and the definitions).

By this chain of arguments, we are able to derive a further link to
understand the underlying action for calculating the expectation value.
In \cite{Wit:89}, Witten constructed a topological quantum field
theory (TQFT) for the Jones polynomial. This theory has its home on
a 3-manifold $\Sigma$ and we will discuss below this 3-manifold.
Let $A$ be a connection of a $SU(2)$ principal bundle over $\Sigma$.
The Chern-Simons action is given by
\begin{equation}
CS(A)=\frac{1}{2\pi}\intop_{\Sigma}tr\left(A\wedge dA+\frac{2}{3}A\wedge A\wedge A\right)\label{eq:CS-action}
\end{equation}
then from \cite{Wit:89} one has the relation
\[
tr(\rho)=V_{L_{\rho}}(i)=\intop DA\,\exp\left(i\cdot CS(A)\right)W_{A}(L_{\rho})
\]
between the trace $tr(\rho)$ and the functional integral over the
action (\ref{eq:CS-action}) where $W_{A}(L)$ is the Wilson loop
along the link $L$ for the connection $A$. With this trick, we get
the action functional (Chern-Simons action) and the observable (Wilson
loop) for the underlying physical theory. The Jones polynomial is
known to be intricately connected with the quantum enveloping algebra
of the Lie algebra of the group $SL(2,\mathbb{C})$, see \cite{ResTur:91}.
In our case, the parameter $q=t=i$ is the 4th root of unity and it
is known that this quantum q-deformation of the Lie algebra $sl_{2}(\mathbb{C})$
yields a finite dimensional modular Hopf algebra. Therefore we have
determined the underlying quantum symmetry (of the initial state at
the Big Bang) as the enveloped algebra $U_{q}(sl_{2}(\mathbb{C}))$.
Furthermore in \cite{Wit:89}, a relation between the $(2+1)-$dimensional
Chern-Simons theory and a $(1+1)-$dimensional conformal field theory
is also discussed. In particular it was shown that the Hilbert space
of pure Chern Simons theories is isomorphic to the space of conformal
blocks of an underlying Conformal Field Theory. This link seem to
imply that there is an underlying $(1+1)-$dimensional theory. We
discussed a similar mechanism in \cite{AsselmeyerMaluga2018d} using
the Morgan-Shalen compactification and will study the relation between
the two approaches in our forthcoming work.

Now we have to determine the 3-manifold $\Sigma$ in the definition
of the Chern-Simons theory. At the first view, we identify $\Sigma$
with the wild 3-sphere. Then this theory is stationary, i.e. it contains
no time variable. But as explained above, the formation of the wild
3-sphere can be seen as a process where the 3-manifold is growing
by attaching 3-dimensional pieces along surfaces. In the definition
of the string algebra, we used Casson handles to define the generators
$e_{i}$. But Casson handles have an inherent 2-dimensional definition
(neighborhood of immersed disks) which is used to define the construction
of the wild 3-sphere (see \cite{AsselmeyerMaluga2016} for a detailed
construction). Then we can see the 3-manifold $\Sigma$ as a non-trivial
cobordism between surfaces (used to define the wild 3-sphere), i.e.
we define the Chern-Simons theory as a $(2+1)-$dimensional theory
right in the sense of Witten \cite{Wit:89}. The 3-manifold is foliated
by the surfaces. To construct this foliation, we introduce light cone
coordinates ($x^{+}=x^{0}+x^{1},x^{-}=x^{0}-x^{1},x^{2}$) together
with the connection 1-form
\[
A(x)=A_{+}(x)dx^{+}+A_{-}(x)dx^{-}+A_{2}(x)dx^{2}.
\]
 (following \cite{Kauffman1998} sec. 4). Now we choose the gauge
$A_{-}=0$ (axial gauge) so that we have a non-zero gauge field for
the future light cone (seen from the Big Bang). Then the Chern-Simons
action simplifies to
\[
CS(A,A_{-}=0)=\frac{1}{2\pi}\intop_{\Sigma}tr\left(A\wedge dA\right)
\]
and the restriction of the $SU(2)$ bundle to the surface leads to
a bundle reduction from $SU(2)$ to $U(1)$ bundle with an abelian
connection $a$ and Chern-Simons form
\[
CS_{U(1)}(a)=\frac{1}{2\pi}\intop_{\Sigma}a\wedge da
\]
This form has a different interpretation in foliation theory: it is
the Godbillon-Vey invariant \cite{GodVey:71}. Recall that a foliation
$(M,F)$ of a manifold $M$ is an integrable subbundle $F\subset TM$
of the tangent bundle $TM$. The leaves $L$ of the foliation $(M,F)$
are the maximal connected submanifolds $L\subset M$ with $T_{x}L=F_{x}\:\forall x\in L$.
A codimension-1 foliation on a 3-manifold $\Sigma$ can be constructed
by a smooth 1-form $\omega$ fulfilling the integrability condition
$d\omega\wedge\omega=0$. Now one defines another one-form $\eta$
by $d\omega=-\eta\wedge\omega$ and the integral over the expression
$gv=\eta\wedge d\eta$ is the Godbillon-Vey invariant. Then the Chern-Simons
invariant in axial gauge defines a codimension-1 foliation of $\Sigma$
where the Chern-Simons invariant is the Godbillon-Vey invariant. The
critical values of the functional $CS_{U(1)}(a)$ are given by $da=0$
and we get a foliation with vanishing Godbillon-Vey invariant. These
foliations are rather trivial (like surface$\times$line or Reeb foliation).
As shown in \cite{HurKat:84,Thu:72}, foliations are really complicate.
In the language of noncommutative geometry, the leaf space of a foliation
with non-vanishing Godbillon-Vey invariants is a von-Neumann algebra
which contains a factor $I\!I\!I$ subalgebra. As shown by Connes
\cite{Connes1984,Connes94}, the Godbillon-Vey class $GV$ can be
expressed as cyclic cohomology class (the so-called flow of weights)
\[
GV_{HC}\in HC^{2}(C_{c}^{\infty}(G))
\]
of the $C^{*}-$algebra for the foliation. Then we define an expression
\[
S=Tr_{\omega}\left(GV_{HC}\right)
\]
uniquely associated to the foliation ($Tr_{\omega}$ is the Dixmier
trace). The expression $S$ generates the action on the factor by
\[
\Delta_{\omega}^{it}=exp(i\,S)
\]
so that $S$ is the action or the Hamiltonian multiplied by the time.
We have evaluated this expression for some cases in \cite{AsselmeyerMaluga2016}
and we interpret it as quantum action. A detailed analysis will be
shifted to our forthcoming work.

But this action is partly satisfactory. In noncommutative geometry,
one introduces a spectral triple with a Dirac operator as main ingredient.
So, let us consider a Dirac operator $D^{\Sigma}$ on $\Sigma$. As
a second ingredient, we introduced a codimension-1 foliation along
the 1-form $a$ which is interpreted as abelian gauge field. To take
this foliation into account, we couple the abelian gauge field $a$
and the spinor $\psi$ to the Dirac-Chern-Simons action functional
on the 3-manifold 
\[
\intop_{\Sigma}\left(\bar{\psi}\,D_{a}^{\Sigma}\psi\,\sqrt{h}d^{3}x+a\wedge da\right)
\]
with the critical points at the solution 
\[
D_{a}^{\Sigma}\psi=0\quad d\eta=\tau(\psi,\psi)
\]
where $\tau(\psi,\psi)$ is the unique quadratic form for the spinors
locally given by $\bar{\psi}\gamma^{\mu}\psi$. Now we consider a
spacetime $\Sigma\times I$, so that the solution is translationally
invariant. Expressed differently, we choose a spacetime with foliation
induced by the foliation of $\Sigma$ extended by a translation. An
alternative description for this choice is by considering the gradient
flow of these equations 
\begin{eqnarray*}
\frac{d}{dt}a & = & da-\tau(\psi,\psi)\\
\frac{d}{dt}\psi & = & D_{a}^{\Sigma}\psi
\end{eqnarray*}
But it is known that this system is equivalent to the Seiberg-Witten
equation for $\Sigma\times I$ by using an appropriate choice of the
$Spin_{C}$ structure \cite{MorSzaTau:96,MorSzaTau:97}. Then this
$Spin_{C}$ structure is directly related to the foliation. Therefore
a non-trivial foliation together with a spectral triple (Dirac operator)
induce a non-trivial solution of the gradient system which results
in a non-trivial solution of the Seiberg-Witten equations. But this
non-trivial solution (i.e. $\psi\not=0,a\not=0$) is a necessary condition
for the existence of an exotic smoothness structure. So we have a
closed circle: we started with a smooth spacetime at the Big Bang
forming the initial state. If this state is a wild 3-sphere, we get
a non-trivial foliation (=non-vanishing Godbillon-Vey invariant) which
produces a non-trivial solution of the Seiberg-Witten equations.

{ Before closing this section, we will discuss the dynamical interpretation
of the string algebra above and the observable. The design $S(Q)$ relative to a Casson handle $Q$ (in our case the unbranched tree) is the sum over all Casson handles leading to the quantum state (the fractal 3-sphere as constructed from $Q$). 
The string algebra for the binary tree (representing the design) is the Clifford algebra of the Hilbert space. From the physics point of view, it is the algebra of fermion field operators. 
Every field operator is given by a path in the binary tree (weighted by some coefficients).
A combination of the results in \cite{AsselmeyerBrans2015} and \cite{AsselmeyerKrol2010} 
showed that the fermion field operators (as elements of the Ocneanu string algebra)
can be also interpreted as the leaf space of a type $I\! I\! I$ foliation (see 
\cite{HurKat:84}) seen as a crossed product of the string algebra and its modular automorphism group. 
This product with the automorphism group is time-dependent representation of the 
field operators (see \cite{ConRov:94}). Therefore, the foliation of type $I\! I\! I$ 
(having a non-zero Godbillon-Vey invariant) is the dynamical interpretation of string algebra.
But we know more because the design was seen as the formation of the fractal 3-sphere as 
given by a sequence of 3-manifolds. This process is given by a sequence of 3-manifold topology change 
which was described in \cite{AsselmeyerKrol2018c}. 
It leads to an inflationary behavior which is approximately described by a de Sitter space
(see \cite{AsselmeyerKrol2018a}).
In \cite{Witten-etal:2022} the algebra of observable for a de Sitter 
space is described to be a von Neumann algebra of type $I\!I_1$. Here we conjecture that 
there must be a relation between our string algebra and this algebra of observables.
}

\section{Conclusion}

{%
In this paper we have worked out a model of the Big Bang driven by
topological considerations. The starting point was the construction
of a spacetime as a global expression of the evolution of the universe.
But the real core of the paper is the construction of the initial
state as a wildly embedded or fractal 3-sphere. Here the construction of a corresponding
operator algebra was the decisive step to understand this state. Here
many interrelations to other theories came to light. Thus, the expectation
value in the operator algebra can always be reinterpreted as a knot
invariant (Jones polynomial). The action of the theory is the Chern-Simons
invariant, which already appears in the description of a (2+1)-dimensional
gravitational theory. In general, these relations to conformal
field and Seiberg-Witten theory are the real strength of this
approach. This work only prepares the ground for further approaches
to the understanding of the initial state at the Big Bang. In
our next work we will interpret and calculate the dark matter density
as a topological quantity.
}
\appendix
\section{Appendix}
{
In this appendix we will describe the methods and results in \cite{AsselmeyerKrol2013} 
to make the paper as self-contained as possible. 
There, we showed that the deformation quantization of a tame embedding is a wild embedding.

At first we start with some definitions. A map $f:N\to M$ between two topological manifolds 
is an embedding if $N$ and $f(N)\subset M$ are homeomorphic to each other. 
An embedding $i:N\hookrightarrow M$ is \emph{tame} if $i(N)$ is represented by a 
finite polyhedron homeomorphic to $N$. Otherwise we call the embedding \emph{wild}.
Let $I:K^{n}\to\mathbb{R}^{n+k}$ be a wild embedding of codimension
$k$ with $k=0,1,2$. Now we assume that the complement
$\mathbb{R}^{n+k}\setminus I(K^{n})$ is non-trivial, 
i.e. $\pi_{1}(\mathbb{R}^{n+k}\setminus I(K^{n}))=\pi\not=1$. 
Wild embedding are usually characterized by this property, but 
if the group $\pi_{1}(\mathbb{R}^{n+k}\setminus I(K^{n}))=1$ is trivial then the
group $\pi_{1}(I(K^{n}))$ must be non-trivial for wild embeddings. 
In section \ref{fractal-3-sphere} we defined the $C^{*}-$algebra $C^{*}(\mathcal{G},\pi$) associated
to the complement $\mathcal{G}=\mathbb{R}^{n+k}\setminus I(K^{n})$
with group $\pi=\pi_{1}(\mathcal{G})$. Therefore the methods of noncommutative
geometry are applicable.\par
For the relation between the tame and wild embedding, we consider the space of geometric
structures on the embedded manifold. In \cite{AsselmeyerKrol2013} we do the 
calculations for Alexanders horned sphere. The space of geometric structures 
with isometry group $SL(2,\mathbb{C})$ admits a Poisson structure. The deformation
quantization of this Poisson structure is known as Drinfeld-Turaev quantization.
In a series of papers it was shown that the deformation quantization 
of the space of geometric structures with isometry group $SL(2,\mathbb{C})$ is the 
Kauffman bracket skein algebra. In case of Alexanders horned sphere, we showed that 
the Kauffman bracket skein algebra is the factor $I\! I_1$ algebra isomorphic to 
the enveloping von Neumann algebra of the $C^*$ algebra defined by the wild embedding. 
In particular for a tame embedding, the skein algebra is trivial (it is only a 1-dimensional
algebra, the center). }
\par

\begin{thebibliography}{999}

\bibitem{PlanckCosmoParameters2013}
P.~{Ade et. al.}
\newblock Planck 2013 results. {XVI}. cosmological parameters.
\newblock arXiv:1303.5076[astro-ph.CO], 2013.

\bibitem{PlanckInflation2013}
P.~{Ade et. al.}
\newblock Planck 2013 results. {XXII}. constraints on inflation.
\newblock arXiv:1303.5082[astro-ph.CO], 2013.

\bibitem{AsselmeyerMaluga2016}
T.~Asselmeyer-Maluga.
\newblock Smooth quantum gravity: {E}xotic smoothness and {Q}uantum gravity.
\newblock In T.~Asselmeyer-Maluga, editor, {\em At the Frontiers of Spacetime:
  Scalar-Tensor Theory, Bell's Inequality, Mach's Principle, Exotic
  Smoothness}. Springer, Switzerland, 2016.
\newblock in honor of Carl Brans's 80th birthday, arXiv:1601.06436.

\bibitem{AsselmeyerMaluga2018d}
T.~Asselmeyer-Maluga.
\newblock Hyperbolic groups, 4-manifolds and quantum gravity.
\newblock {\em Journal of Physics: Conference Series}, {\bf 1194}:012009, 2019.
\newblock arXiv: 1811.04464.

\bibitem{AsselmeyerMaluga2019}
T.~Asselmeyer-Maluga.
\newblock Braids, 3-Manifolds, Elementary Particles: Number Theory and Symmetry in Particle Physics.
\newblock {\em Symmetry}, {\bf 11}:1298, 2019. 
\newblock doi:10.3390/sym11101298; arXiv: 1910.09966.

\bibitem{AssBra:08}
T.~Asselmeyer-Maluga and C.H. Brans.
\newblock {\em Exotic Smoothness and Physics}.
\newblock World Scientific, Singapore, 2008.

\bibitem{AsselmeyerBrans2015}
T.~Asselmeyer-Maluga and C.H. Brans.
\newblock How to include fermions into general relativity by exotic smoothness.
\newblock {\em Gen. Relativ. Grav.}, {\bf 47}:30, 2015.
\newblock DOI 10.1007/s10714-015-1872-x, arXiv: 1502.02087.

\bibitem{AsselmeyerKrol2010}
T.~Asselmeyer-Maluga and J.~Kr{\'o}l.
\newblock Exotic smooth {${\mathbb{R}}^4$}, noncommutative algebras and quantization.
\newblock arXiv:1001.0882.

\bibitem{AsselmeyerKrol2012}
T.~Asselmeyer-Maluga and J.~Kr{\'o}l.
\newblock On topological restrictions of the spacetime in cosmology.
\newblock {\em Mod. Phys. Lett. A}, {\bf 27}:1250135, 2012.
\newblock arXiv:1206.4796.

\bibitem{AsselmeyerKrol2013c}
T.~Asselmeyer-Maluga and J.~Kr{\'o}l.
\newblock Decoherence in quantum cosmology and the cosmological constant.
\newblock {\em Mod. Phys. Lett. A}, {\bf 28}(34):1350158, 2013.
\newblock arXiv:1309.7206.

\bibitem{AsselmeyerKrol2013}
T.~Asselmeyer-Maluga and J.~Kr{\'o}l.
\newblock Quantum geometry and wild embeddings as quantum states.
\newblock {\em Int. J. of Geometric Methods in Modern Physics}, {\bf
  10}(10):1350055, 2013.
\newblock arXiv:1211.3012.

\bibitem{AsselmeyerKrol2018a}
T.~Asselmeyer-Maluga and J.~Krol.
\newblock How to obtain a cosmological constant from small exotic {${\mathbb
  R}^4$}.
\newblock {\em Physics of the Dark Universe}, {\bf 19}:66--77, 2018.
\newblock arXiv:1709.03314.

\bibitem{AsselmeyerKrol2018c}
T.~Asselmeyer-Maluga and J.~Krol.
\newblock A topological model for inflation.
\newblock arXiv:1812.08158.

\bibitem{AsselmeyerKrol2018b}
T.~Asselmeyer-Maluga and J.~Krol.
\newblock A topological approach to neutrino masses by using exotic smoothness.
\newblock {\em Mod. Phys. Lett. A}, {\bf 34}:1950097, 2019.
\newblock DOI: 10.1142/S0217732319500974 , arXiv:1801.10419.

\bibitem{AsselmeyerRose2012}
T.~Asselmeyer-Maluga and H.~Ros{\'e}.
\newblock On the geometrization of matter by exotic smoothness.
\newblock {\em Gen. Rel. Grav.}, {\bf 44}:2825 -- 2856, 2012.
\newblock DOI: 10.1007/s10714-012-1419-3, arXiv:1006.2230.

\bibitem{Biz:95}
Z.~Bizaca.
\newblock An explicit family of exotic {C}asson handles.
\newblock {\em Proc. AMS}, 123:1297--1302, 1995.

\bibitem{BizGom:96}
{\u Z}.~Bi{\u z}aca and R~Gompf.
\newblock Elliptic surfaces and some simple exotic {${\Bbb {R}}^4$}'s.
\newblock {\em J. Diff. Geom.}, {\bf 43}:458--504, 1996.


\bibitem{Borchers2000}
H.J. Borchers.
\newblock On revolutionizing quantum field theory with {T}omita's modular
  theory.
\newblock {\em J. Math. Phys.}, 41:3604 -- 3673, 2000.

\bibitem{Bra:94b}
C.H. Brans.
\newblock Exotic smoothness and physics.
\newblock {\em J. Math. Phys.}, {\bf 35}:5494--5506, 1994.

\bibitem{Bra:94a}
C.H. Brans.
\newblock Localized exotic smoothness.
\newblock {\em Class. Quant. Grav.}, {\bf 11}:1785--1792, 1994.

\bibitem{Witten-etal:2022}
V.~Chandrasekaran, R.~Longo, G.~Penington, and E.~Witten
\newblock An Algebra of Observables for de Sitter Space.
\newblock arXiv:2206.10780.

\bibitem{Connes1984}
A.~Connes.
\newblock A survey of foliations and operator algebras.
\newblock {\em Proc. Symp. Pure Math.}, 38:521--628, 1984.
\newblock see www.alainconnes.org.

\bibitem{Connes94}
A.~Connes.
\newblock {\em Non-commutative geometry}.
\newblock Academic Press, 1994.

\bibitem{ConRov:94}
A.~Connes and C.~Rovelli.
\newblock Von {N}eumann algebra automorphisms and time-thermodynamics relation in
  generally covariant quantum theories.
\newblock {\em Class. Quan. Grav.}, {\bf 11}(12):2899 -- 2917, 1994.

\bibitem{Don:83}
S.~Donaldson.
\newblock An application of gauge theory to the topology of 4-manifolds.
\newblock {\em J. Diff. Geom.}, {\bf 18}:269--316, 1983.

\bibitem{FreQui:90}
M.~Freedman and F.~Quinn.
\newblock {\em Topology of 4-{M}anifolds}.
\newblock Princeton Mathematical Series. Princeton University Press, Princeton,
  1990.

\bibitem{Fre:82}
M.H. Freedman.
\newblock The topology of four-dimensional manifolds.
\newblock {\em J. Diff. Geom.}, {\bf 17}:357 -- 454, 1982.

\bibitem{Fre:83}
M.H. Freedman.
\newblock The disk problem for four-dimensional manifolds.
\newblock In {\em Proc. Internat. Cong. Math. Warzawa}, volume~{\bf 17}, pages
  647 -- 663, 1983.

\bibitem{GodVey:71}
C.~Godbillon and J.~Vey.
\newblock Un invariant des feuilletages de codimension.
\newblock {\em C. R. Acad. Sci. Paris Ser. A-B}, 273:A92, 1971.

\bibitem{GomSti:1999}
R.E. Gompf and A.I. Stipsicz.
\newblock {\em 4-manifolds and {K}irby {C}alculus}.
\newblock American Mathematical Society, 1999.

\bibitem{GoHaJo:89}
F.~Goodman, P.~de~la Harpe, and V.~Jones.
\newblock {\em Coxeter graphs and towers of algebras}.
\newblock Springer, {MSRI} publications edition, volume~{\bf 14}, 1989.

\bibitem{HawEll:94}
S.W Hawking and G.F.R. Ellis.
\newblock {\em The Large Scale Structure of Space-Time}.
\newblock Cambridge University Press, 1994.

\bibitem{HawHar:83}
J.B. Hartle and S.W. Hawking.
\newblock Wave function of the universe.
\newblock {\em Phys. Rev. D}, {\bf 28}:2960, 1983.
\newblock http://dx.doi.org/10.1103/PhysRevD.28.2960.

\bibitem{HawTur:98a}
S.W. Hawking and N.~Turok.
\newblock Open inflation without false vacua.
\newblock {\em Phys. Lett.}, B425:25--32, 1998.

\bibitem{Hein2018}
H.-J.~Hein, S.~Sun, J.~Viaclovsky, and R.~Zhang.
\newblock Nilpotent structures and collapsing Ricci-flat metrics on K3 surfaces.
\newblock arXiv:1807.09367.

\bibitem{Hein2021}
H.-J.~Hein, S.~Sun, J.~Viaclovsky, and R.~Zhang.
\newblock Gravitational instantons and del Pezzo surfaces.
\newblock arXiv:2111.09287.

\bibitem{HurKat:84}
S.~Hurder and A.~Katok.
\newblock Secondary classes and transverse measure theory of a foliation.
\newblock {\em BAMS}, 11:347 -- 349, 1984.
\newblock announced results only.

\bibitem{Jon:83}
V.~Jones.
\newblock Index of subfactors.
\newblock {\em Invent. Math.}, {\bf 72}:1--25, 1983.

\bibitem{Jon:85}
V.F.R. Jones.
\newblock A polynomial invariant for knots via von {N}eumann algebras.
\newblock {\em BAMS}, {\bf 12}(1):103, 1985.

\bibitem{Kauffman1998}
L.H. Kauffman.
\newblock Functional integration and the {K}ontsevich integral.
\newblock In {\em Yang-Baxter Systems, Nonlinear Models and Their
  Applications}, Proceedings of the APCTP-Nankai Symposium, Singapore, 1999.
  World Scientific.
\newblock https://doi.org/10.1142/4271, math/9811137.

\bibitem{Lebrun96}
C.~LeBrun.
\newblock Four-manifolds without {E}instein metrics.
\newblock {\em Math. Res. Lett.}, 3:133--147, 1996.

\bibitem{MorSzaTau:96}
J.W. Morgan, Z.~Szabo, and C.H. Taubes.
\newblock A product formula for the {S}eiberg-{W}itten invariants and the
  generalized {T}hom conjecture.
\newblock {\em J. Diff. Geom.}, {\bf 44}:706--788, 1996.

\bibitem{MorSzaTau:97}
J.W. Morgan, Z.~Szabo, and C.H. Taubes.
\newblock Product formulas along $t^3$ for {S}eiberg-{W}itten invariants.
\newblock {\em Mathematical Research Letters}, {\bf 4}:915--929, 1997.

\bibitem{Mur:86}
H.~Murakami.
\newblock A recursive calculation of the {A}rf invariant of a link.
\newblock {\em J. Math. Soc. Japan}, {\bf 38}(2):335, 1986.

\bibitem{Ocn:88}
A.~Ocneanu.
\newblock Quantized groups, string algebras and {G}alois theory for algebras.
\newblock In Evans and Takesaki, editors, {\em Operator Algebras and
  Applications}, pages 119--172, 1988.

\bibitem{ResTur:91}
N.Y. Reshetikhin and V.~Turaev.
\newblock Invariants of three-manifolds via link polynomials and quantum
  groups.
\newblock {\em Inv. Math.}, {\bf 103}:547--597, 1991.

\bibitem{Thu:72}
W.~Thurston.
\newblock Noncobordant foliations of {$S^3$}.
\newblock {\em BAMS}, 78:511 -- 514, 1972.

\bibitem{Wit:89.2}
E.~Witten.
\newblock 2+1 dimensional gravity as an exactly soluble system.
\newblock {\em Nucl. Phys.}, {\bf B311}:46--78, 1988/89.

\bibitem{Wit:89}
E.~Witten.
\newblock Quantum field theory and the {J}ones polynomial.
\newblock {\em Commun. Math. Phys.}, {\bf 121}:351--400, 1989.

\bibitem{Yau:1978}
S.-T. Yau.
\newblock On the {R}icci curvature of a compact {K}\"ahler manifold and the complex
  {M}onge-{A}mp\`ere equation.
\newblock {\em Communications on Pure and Applied Mathematics}, {\bf
  31}:339--411, 1978.

\end{thebibliography}
\section*{Acknowledgments}
We acknowledge the remarks and questions of the referees, 
improving the argumentation in the paper and increasing the readability.

\end{document}